# Record-high Mobility and Extreme Magnetoresistance on Kagome-lattice in Compensated Semimetal $Ni_3In_2S_2$

Hongwei Fang[1,2†], Meng Lyu[3†], Hao Su[1,2], Jian Yuan[1,2], Yiwei Li[1], Lixuan Xu[4], Shuai Liu[1,2], Liyang Wei[1,2], Xinqi Liu[1,5], Haifeng Yang[1], Qi Yao[1,5], Meixiao Wang[1,5], Yanfeng Guo[1], Wujun Shi[6,7*], Yulin Chen[1,5,8*], Enke Liu[3*] and Zhongkai Liu[1,5*]

[1]School of Physical Science and Technology, ShanghaiTech University, Shanghai 201210, China

[2]University of Chinese Academy of Sciences, Beijing 100049, China

[3]Beijing National Laboratory for Condensed Matter Physics, Institute of Physics, Chinese Academy of Sciences, Beijing 100190, China

[4]State Key Laboratory of Low Dimensional Quantum Physics, Department of Physics, Tsinghua University, Beijing 100084, China

[5]ShanghaiTech Laboratory for Topological Physics, ShanghaiTech University, Shanghai 201210, China

[6]Center for Transformative Science, ShanghaiTech University, Shanghai, 201210, China

[7]Shanghai High Repetition Rate XFEL and Extreme Light Facility (SHINE), ShanghaiTech University, Shanghai, 201210, China

[8]Department of Physics, University of Oxford, Oxford OX1 3PU, United Kingdom

†These authors contributed equally to this work.

*Corresponding authors (emails: shiwujun@shanghaitech.edu.cn (Shi W); yulin.chen@physics.ox.ac.uk (Chen Y); ekliu@iphy.ac.cn (Liu E); liuzhk@shanghaitech.edu.cn (Liu Z))

## ABSTRACT

**The kagome-lattice crystal hosts various intriguing properties including the frustrated magnetism, charge order, topological state, superconductivity and correlated phenomena. To achieve high-performance kagome-lattice compounds**

**for electronic and spintronic applications, careful tuning of the band structure would be desired. Here, the electronic structures of kagome-lattice crystal $Ni_3In_2S_2$ were investigated by transport measurements, angle-resolved photoemission spectroscopy as well as *ab initio* calculations. The transport measurements reveal $Ni_3In_2S_2$ as a compensated semimetal with record-high carrier mobility (~8683 $cm^2$ $V^{-1}$ $S^{-1}$ and 7356 $cm^2$ $V^{-1}$ $S^{-1}$ for holes and electrons) and extreme magnetoresistance (15518% at 2 K and 13 T) among kagome-lattice materials. These extraordinary properties are well explained by its band structure with indirect gap, small electron/hole pockets and large bandwidth of the 3*d* electrons of Ni on the kagome lattice. This work demonstrates that the crystal field and doping serve as the key tuning knobs to optimize the transport properties in kagome-lattice crystals. Our work provides material basis and optimization routes for kagome-lattice semimetals as electronics and spintronics applications.**

## INTRODUCTION

In recent years, the kagome-lattice crystals have attracted great research interests due to its unique magnetic and electronic properties[1-5]. The kagome-lattice crystals are ideal platform to seek for the quantum spin liquid ground states due to the large geometric frustrations[1, 6-10]; the kagome-lattice could host Dirac cone-type dispersions similar to the honeycomb lattice[11-13], flat bands due to the completely destructive interference of Bloch wave functions[11, 13, 14], topological electronic states (*e.g.*, Weyl cones and topological surfaces states) as well as electronic instabilities such as charge density waves and superconductivity[15-19]. These unique properties render kagome-lattice compounds fascinating candidates for electronic and spintronic applications.

For device applications, high performance kagome-lattice compounds (such as high carrier mobility, robust magnetism and high $T$c superconductivity) are highly desired[20-23]. As an example, kagome-lattice naturally hosts fast-moving Dirac fermions which are feasible for high mobility electronic devices[12-14]. However, the typical mobility of existing kagome-lattice crystals is ~100 to 1000 $cm^2\ V^{-1}s^{-1}$[20, 24, 25], much smaller than the typical Dirac electron systems (such as graphene, ~$10^4$ $cm^2\ V^{-1}s^{-1}$)[26-28] and high performance MOSFET materials (*e.g.*, GaAs, ~$10^4$ $cm^2\ V^{-1}\ s^{-1}$)[29-31]. Therefore, careful material optimization and tuning of the band structure would be a necessity to achieve kagome-lattice crystals with superior transport properties and facilitate kagome-lattice materials applications in electronic and spintronic devices.

Among the kagome-lattice compounds, the great tunability of the 3*d* transition metal intermetallic compounds provides a versatile platform for the search and optimization of the physical properties in kagome-lattice crystals, including FeSn, CoSn, $YMn_6Sn_6$ *etc*[11, 13, 14]. In this work, we demonstrate such an effort in the kagome-lattice shandite $Ni_3In_2S_2$, a non-magnetic isostructural counterpart of the recently discovered magnetic topological Weyl semimetal $Co_3Sn_2S_2$[15, 16, 32, 33].

With Ni substituting Co and In for Sn, we effectively tailor the band structure and achieve record-high mobility (~8683 $cm^2 V^{-1} S^{-1}$ and 7356 $cm^2 V^{-1} S^{-1}$ for holes and electrons) and electron-hole compensated extremely large and unsaturated magnetoresistance (15518% at 2 K and 13 T) among all the existing kagome-lattice crystals. Via band structure investigation and analysis using angle resolved photoemission spectroscopy (ARPES) and *ab initio* calculations, we attribute these superior transport properties to a) the delicate band arrangement under the crystal field which forms a bandgap via the topological phase transition ($Co_3Sn_2S_2$ possesses inverted 3*d* bands) and restores the bandwidth of the Ni 3*d* electrons on the kagome-lattice and b) proper doping which forms small and compensated electron and hole pockets at the $E_F$. Our results demonstrate a compensated semimetal $Ni_3In_2S_2$ with high mobility and extreme magnetoresistance (XMR) and illustrate possible route in the modification of the electronic structure of kagome-lattice compounds in achieving superior transport properties, which demonstrate possible routes into utilizing kagome-lattice materials as high mobility electronic and spintronic applications.

## EXPERIMENTAL SECTION

### Crystal growth:

Single crystals of $Ni_3In_2S_2$ were grown through the solid state chemical reaction route. Mixtures of high-purity elements Ni (Macklin, 99.99%), In (Aladdin, 99.99%) and Sulfur (Adamas, 99.999%) in stoichiometric ratio were put into an alumina crucible and sealed inside an evacuated quartz tube. The assembly was heated up to 1000℃ within 20 hours in the furnace and held for 30 hours. Then it is slowly cooled at rate of 2°C/h to 500°C and switched off the furnace to cooled down to room temperature.

**Electrical transport measurements:**

The electrical transport measurements were carried out in the physical property measurement system (PPMS, 14T) between 2K and room temperature, using a sample of typical dimension $0.1\times 0.5 \times 1.5$ mm$^3$. Standard four-probe method was applied for the longitudinal resistivity and the Hall Effect measurements with a current along *a*-axis and magnetic fields parallel to *c*-axis. To eliminate the influence of misalignment of the lead contact, all magnetoresistance and Hall-effect measurements were conducted by scanning both negative and positive magnetic fields.

**Angle-resolved photoemission spectroscopy:**

ARPES measurements were performed at the beamline I05 of the Diamond Light Source (DLS) with Scienta R4000 analyzer and beamline BL03U of Shanghai Synchrotron Radiation Facility (SSRF) with Scienta DA30 analyzer. The photon-energy ranges of data acquisition for DLS and SSRF are 52–200 eV and 58–114 eV, respectively. The samples were cleaved in situ at 23 K and measured in ultrahigh vacuum with a base pressure of better than $5 \times 10^{-11}$ Torr. The energy and momentum resolution were 10 meV and 0.2°, respectively.

**Theoretical calculation:**

The first-principles calculations were performed using the Vienna ab initio Simulation

Package (VASP)[34]. The interactions between the valence electrons and ion cores are described by the projector augmented wave method[35, 36], and exchange-correlation potential is formulated by the generalized gradient approximation with the Perdew-Burke-Ernzerhof (PBE) scheme[34]. The Γ-centered 10×10×10 k points are used for the first Brillouin-zone sampling. The spin-orbit coupling (SOC) is included in all the calculations. The tight-binding Hamiltonian was constructed using the maximally localized Wannier functions which was provided by Wannier90 package[37]. The surface states were calculated by the surface Green's function method[38] based the tight-binding Hamiltonian. The experiment lattice constant (Inorganic Crystal Structure Database no. 415258) was used in the calculations.

## RESULTS AND DISCUSSION

First, we characterized the basic properties of $Ni_3In_2S_2$. $Ni_3In_2S_2$ has a rhombohedral lattice structure with space group $R\bar{3}m$ (No.166), The conventional cell and primitive cell are shown in Fig. 1a with the conventional lattice constants to be $a$ = $b$ = 5.37 Å, $c$ = 13.56 Å. The crystal is formed by sequenced In-[S-(Ni3-In)-S] layers along the *c* direction where Ni atoms form a kagome-lattice (Fig. 1b) sandwiched between two hexagonal S atoms. The typical samples for our measurement have sizes around several millimeters and the high quality of the single-crystalline samples used in this work is demonstrated by the single-crystal X-ray diffraction angle scan (Fig. 1c). Fig. 1d presents the temperature dependent longitudinal resistivity $\rho_{xx}$(T) under different magnetic fields. Upon cooling from room temperature to 2 K, the zero-field $\rho_{xx}$(T) continuously decreases and then flattens at low temperature, without any signature of phase transition, suggesting the absence of long-range magnetic order in $Ni_3In_2S_2$. Both the quite low residual resistivity of 0.108 $\mu\Omega$ cm and the large residual resistance ratio (RRR) of 215 reflect the very high quality of the studied crystal. Upon the increasing field, we noticed a significant upturn of the resistivity at low temperatures, which indicates a large magnetoresistance (MR) effect of $Ni_3In_2S_2$, and further, a semi-metallic with small Fermi pockets. The similar low-temperature resistivity upturn behavior has also been observed in other semi-metallic compounds, such as TaAs, $PtBi_2$, and $WP_2$[39-42]. In addition, the calculated three-dimensional Fermi surface were shown in Fig. 1e-f. Electron pockets near Γ point and hole pockets near W points were observed and the electron pockets and hole pockets possess similar volumes, demonstrating $Ni_3In_2S_2$ as a nearly compensated semimetal.

The above analysis hints the similar transport properties as nearly compensated semimetal and encourages us to explore the magneto-transport properties of $Ni_3In_2S_2$, which is summarized in Fig. 2. Fig. 2a shows the magnetic-field dependent MR for $Ni_3In_2S_2$ single crystal at different temperatures. The MR is greatly enhanced with applied external magnetic field along the $c$-axis direction, displaying no signature of saturation and reaching a high value of 15518% at 2 K and 13 T, representing an XMR effect in a semimetal. This XMR effect is rarely observed in other kagome-lattice materials, suggesting the unique magneto-transport property in $Ni_3In_2S_2$. Upon raising the temperature, the MR decreases dramatically and becomes negligible at 50 K. In the inset of Fig. 2a, we fitted the MR at T = 2 K by $AH^n$ where $n$ is estimated to be 1.9, which further indicates $Ni_3In_2S_2$ may be a nearly compensated semimetal with high carrier mobility (the perfectly compensated semimetal with equal density of electron ($n_e$) and hole ($n_h$) type carriers gives $MR = u_e u_h H^2$, where $u_e/u_h$ are electron/hole mobility, respectively).

Furthermore, to evaluate the carrier-related parameters, we estimated the carrier density and mobility from the Hall resistivity ($\rho_{yx}$) and magneto-resistivity ($\rho_{xx}$) at different temperatures. The $\rho_{yx}$ at high temperatures is almost linear and positive, implying that the majority carrier is hole type (see SI). Fig. 2b plots $\rho_{yx}$ versus magnetic field at various temperatures, which exhibits a nonlinear behavior and persists down to 2 K, reflecting typical characteristic of multi-type carriers. We thus extracted the carrier density and mobility of $Ni_3In_2S_2$ at low temperatures using two-band model, by simultaneously fitting

$$\rho_{yx} = \frac{B}{e}\frac{(n_h\mu_h^2 - n_e\mu_e^2) + (n_h - n_e)\mu_h^2\mu_e^2B^2}{(n_h\mu_h + n_e\mu_e)^2 + (n_h - n_e)^2\mu_h^2\mu_e^2B^2}$$

$$\rho_{xx} = \frac{1}{e}\frac{(n_h\mu_h + n_e\mu_e) + (n_h\mu_e + n_e\mu_h)\mu_h\mu_eB^2}{(n_h\mu_h + n_e\mu_e)^2 + (n_h - n_e)^2\mu_h^2\mu_e^2B^2}$$

, where $n_h(n_e)$ and $u_h(u_e)$ are the hole (electron) density and mobility, respectively. Fig. 2c and Fig. 2d displays the fitting results of the mobility of the carriers and carrier density at different temperatures. At 2 K, $n_h$=2.49×10$^{21}$ cm$^{-3}$ and $n_e$ =2.42×10$^{21}$ cm$^{-3}$ are quite close to each other, and are highly consistent with the results obtained by band calculations, which gives $n_h$=1.37×10$^{21}$ cm$^{-3}$ and $n_e$=1.36×10$^{21}$ cm$^{-3}$, respectively (the difference may be due to simplified nature of the two-band model). These results indicate $Ni_3In_2S_2$ is indeed a compensated semimetal. Besides, the holes and electrons mobility at 2 K are 8683 cm$^2$ V$^{-1}$ s$^{-1}$ and 7356 cm$^2$ V$^{-1}$ s$^{-1}$, respectively, which are the highest values among all reported kagome-lattice materials (see Fig. 2e)[20, 43]. The large carrier mobility and compensated carrier concentration in turn explains the above mentioned XMR effect in this system, in the similar mechanism as proposed in compensated semimetals[44]. Interestingly, we note the mobility in $Ni_3In_2S_2$ is multiple times higher than that in the isostructural $Co_3Sn_2S_2$[20], suggesting the role of different crystal field and doping in tuning the band structure and the transport properties, as elaborated later. These excellent transport properties observed in $Ni_3In_2S_2$ could provide an ideal platform for studying the electronic transport behavior for advanced electronic or spintronic devices based on kagome physics.

In order to trace the origin of the high mobility and non-saturated XMR effect, we systematically investigated the electronic structure of $Ni_3In_2S_2$ by ARPES. The three-dimensional Brillouin zone (BZ) of the primitive cell and the projected surface BZ of the conventional cell in the (001) plane is shown in Fig. 3a, with the momentum axis labelled. After cleaving, flat and shiny surface was created, ideal for ARPES measurement, the high crystal quality was confirmed by the Laue pattern and the topography image of the cleaved surface measured by scanning tunneling microscope (STM) (see in Fig. 3b), which confirmed the cleavage surface as the (001) surface. The measured high symmetry dispersion along the $\bar{K}-\bar{\Gamma}-\bar{K}$ and $\bar{M}-\bar{\Gamma}-\bar{M}$ directions are shown in Fig. 3c(i) and d(i). Due to significant $k_z$ broadening effect observed (see SI for the photon energy dependent ARPES measurement), ARPES data captures both dispersions from Γ and T point. Therefore, hole pockets from W and electronic pockets from Γ (labeled as BVB/BCB in Fig. 3c(ii), see the calculated band structure in the primitive cell in Fig. 3e) are observed near $\bar{K}$and $\bar{\Gamma}$, respectively, showing excellent agreement with the projected band structure from the slab calculations (Fig. 3c, d). The constant energy contours (CECs) further reveal clear hole pockets near $\bar{\mathrm{K}}$ and electron pocket near $\bar{\Gamma}$ (Fig. 3f), identical to the calculation results in Fig. 3g (see detailed comparison in Fig. S3, which gives the estimate of three-dimensional hole/electron concentration as $n_h = 1.37\times10^{21}$ cm$^{-3}$ and $n_e = 1.36\times10^{21}$ cm$^{-3}$, respectively, further proving the electron-hole compensation nature in $Ni_3In_2S_2$). We also note the sharp surface states (SS) near the $\bar{M}$ point could be identified, contributing to the parallel lines near the $\bar{M}$ point (the SS are marked by the red arrows, see SI for detail). The excellent agreement between experiments and calculations proves the validity of the calculation. Combined with transport and calculation results of the electron and hole concentrations, we further confirmed the nearly compensated semimetal nature in $Ni_3In_2S_2$, which are key to the high mobility and non-saturated XMR effect in $Ni_3In_2S_2$.

We interpret the superior transport behavior based on the calculated band structure as presented in Fig. 3e. The orbital analysis suggests the electron pockets near Γ and hole pockets near W are originated from the Ni 3*d* orbitals on the kagome-lattice, which indicates that the excellent transport properties are closely related to the kagome structure. To further elaborate the relation between the unique electronic structure and the transport properties in $Ni_3In_2S_2$, we further performed systematic calculations on the band structure of four isostructural compounds: $Co_3Sn_2S_2$, $Ni_3Sn_2S_2$, $Ni_3In_2S_2$ and $Co_3In_2S_2$ and explore their band evolution to uncover the origin of the superior transport properties in $Ni_3In_2S_2$. As Fig. 4a-d shows, in $Co_3Sn_2S_2$, $Ni_3Sn_2S_2$, and $Ni_3In_2S_2$, the $dz^2$ electron-like bands and $dx^2$-$y^2$ hole-like bands are cutting through $E_F$ while the $dz^2$ hole-like bands dominate in $Co_3In_2S_2$ due to the lack of electrons. Due to the different crystal field and spin orbit coupling strength, there is clear band inversion between $dx^2$-$y^2$ and $dz^2$, leading to the topological Weyl semimetal phase in the ferromagnetic $Co_3Sn_2S_2$[32] and topological insulator phase in the paramagnetic $Ni_3Sn_2S_2$. The inverted band structure creates local band gaps and reduces the band width of the $dx^2$-$y^2$ and $dz^2$ bands. In $Ni_3In_2S_2$ and $Co_3In_2S_2$, the band inversion between $dx^2$-$y^2$ and $dz^2$ was cancelled and their bandwidth restored, leading to an indirect semiconductor-like gap around the Fermi level and allowing fast moving carriers from the spherical Fermi surfaces in indirect gaps (Fig. 4e). These characteristic carriers bring about high mobility, long mean free path, and XMR effect. Meanwhile, the different valence electrons in Co/Ni and In/Sn tunes the Fermi level and controls the carrier concentration. In $Co_3Sn_2S_2$, $Ni_3Sn_2S_2$ and $Ni_3In_2S_2$, the calculated concentrations of electron/hole carriers are less than the $2\times10^{21}$ $cm^{-3}$ and almost compensated, while in $Co_3In_2S_2$ the hole type carrier dominates and has concentration of $8\times10^{21}$ $cm^{-3}$ (Fig. 4f). Combining the bandwidth and $E_F$ position together, $Ni_3In_2S_2$ is the optimal system among the family with large bandwidth, small carrier concentration and compensated carrier, which explains its excellent high mobility and non-saturated XMR effect.

## CONCLUSIONS

In summary, we have investigated the kagome-lattice material $Ni_3In_2S_2$ by magneto-transport measurement and electronic structure analysis. Our results reveal the high-mobility and non-saturated XMR in this compound which could be explained by the large bandwidth and electronic structure in compensated semimetal with an indirect gap. Such superior property could be attributed to the crystal field and spin-orbit interaction strength which controls the bandwidth via the topological phase transition; as well as the chemical doping which tunes the carrier type and concentration. Our results illustrate the key tuning knob of the electronic structure and key transport properties in the kagome-lattice crystals. The $(Co,Ni)_3(Sn,In)_2S_2$ provides an ideal platform to investigate magnetism and topological property, as well as achieve high mobility electronic and spintronic applications in kagome-lattice materials.

## ACKNOWLEDGEMENTS

This work is supported by grant from the National Key R&D program of China (Grant Nos. 2017YFA0305400 and 2019YFA0704900) and Chinese Academy of Science-Shanghai Science Research Center (Grant No. CAS-SSRC-YH-2015-01). Y. C. acknowledges the support from the Engineering and Physical Sciences Research Council Platform Grant (Grant No. EP/M020517/1). The Major Research Plan of the National Natural Science Foundation of China (NSFC, Grant No. 92065201 to Y.G.), Shanghai Municipal Science and Technology Major Project (Grant No. 2018SHZDZX02 to Y. C. and Z. L.). E. L. acknowledges the support from NSFC (Grant Nos. 52088101 and 11974394), and the Strategic Priority Research Program (B) of the Chinese Academy of Sciences (Grant No. XDB33000000). W. Shi acknowledges the support from Shanghai Committee of Science and Technology (22ZR1441800), Shanghai-XFEL Beamline Project (SBP) (31011505505885920161A2101001). Y. L acknowledges the support from the National Natural Science Foundation of China (12104304). The calculations were carried out at the Scientific Data Analysis Platform of Center for Transformative Science and the HPC Platform of ShanghaiTech University Library and Information Services. The synchrotron based ARPES measurement was conducted in Beamline I05 of Diamond Light Source, and BL03U of Shanghai Synchrotron Radiation Facility. We acknowledge the Analytical Instrumentation Center of ShanghaiTech University for Laue diffraction measurements.

## AUTHOR CONTRIBUTIONS

Liu Z and Chen Y conceived the project; Fang H, Li Y, Xu L and Lyu M performed the ARPES, XRD and electron transport study with the help from Yang H and Liu E; Shi W performed the theoretical calculation; Liu S and Wei L performed the STM study with the help from Wang M; Su H and Yuan J synthesized the crystals; Liu X and Yao Q performed the literature research. All authors contributed to the general discussion.

## CONFLICT OF INTEREST

The authors declare that they no conflict of interest.

## SUPPORTING INFORMATION

Supporting data are available in the online version of the paper.

## Figures

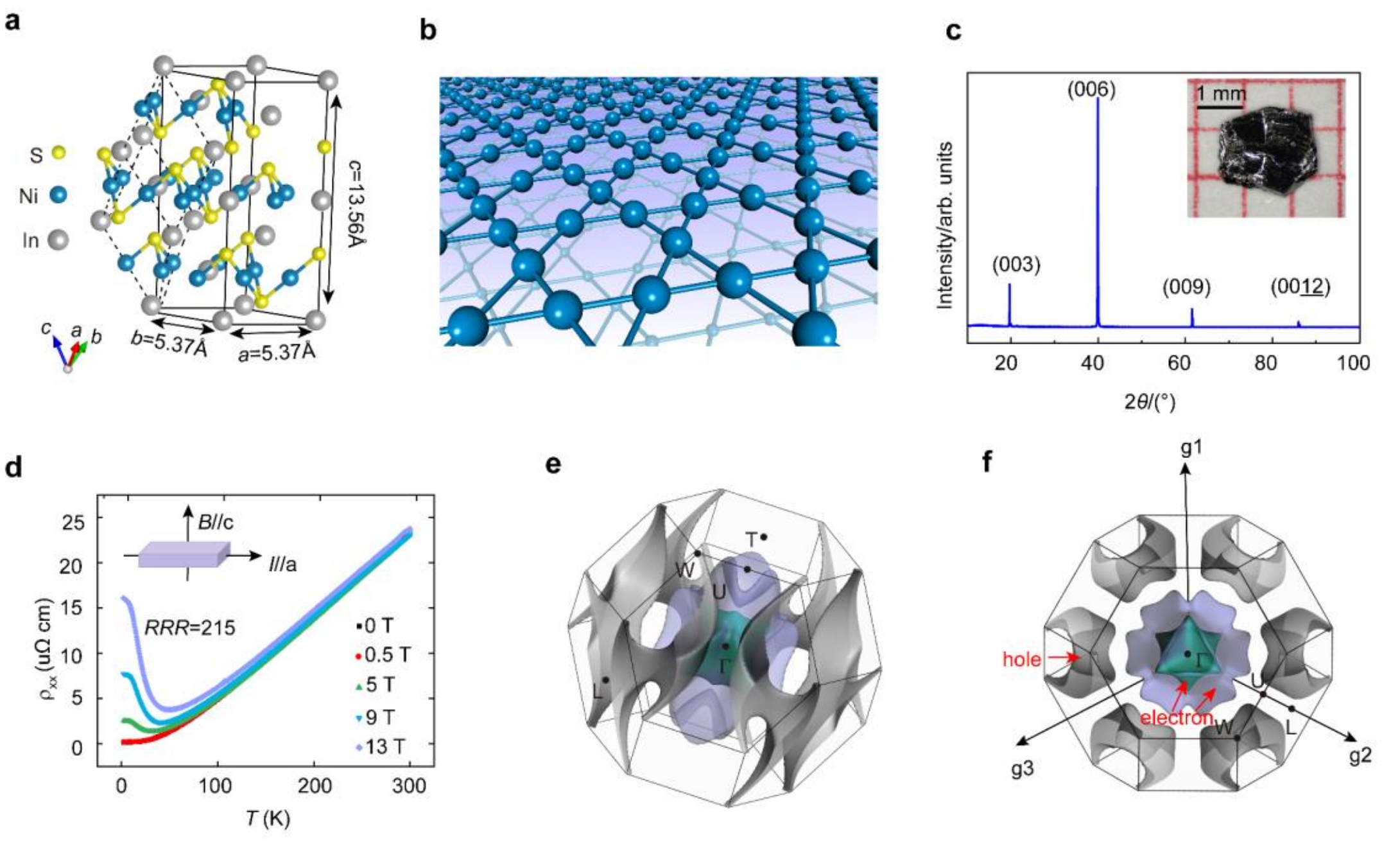

**Figure 1** Basic characterization of $Ni_3In_2S_2$. (a) Conventional (solid lines) and primitive (dotted lines) cell of $Ni_3In_2S_2$. (b) Illustration of the kagome-lattice formed by Ni. (c) XRD pattern of $Ni_3In_2S_2$ measured at room temperature. Inset: Photograph of the high-quality $Ni_3In_2S_2$ single crystal. (d) Temperature dependent resistivity under different magnetic fields. Inset: configuration of the applied electrical current and magnetic fields. (e) Three-dimensional map and (f) top view of the calculated Fermi surface. Purple and green sheets represent electron pockets and gray sheets represent hole pockets.

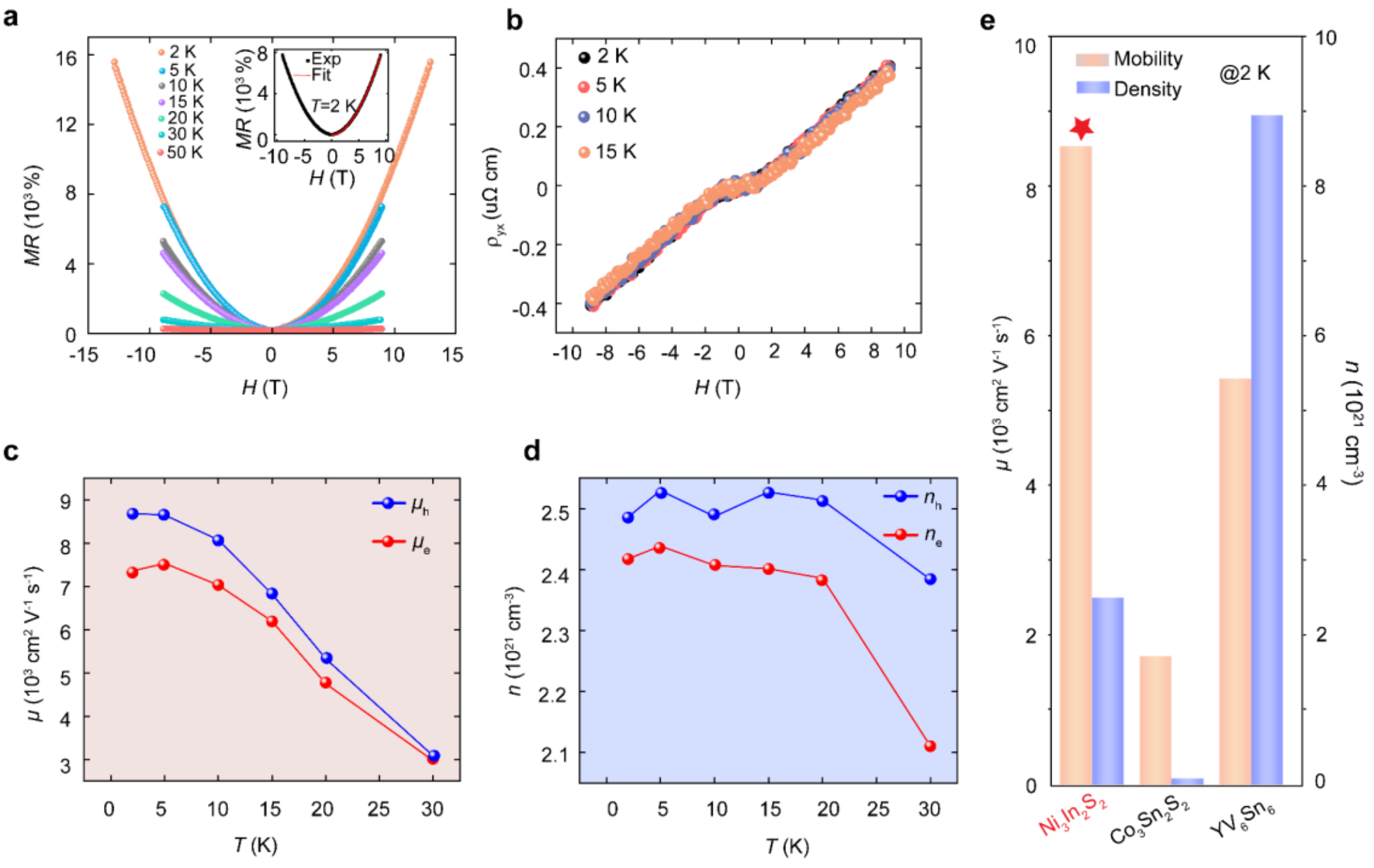

**Figure 2** Magnetic transport measurement of $Ni_3In_2S_2$. (a) Magnetoresistance as a function of the magnetic field at 2–50 K. In the inset, the experimental data at 2 K have been fitted using $MR = AH^n$, yielding $n$ = 1.9. (b) Hall resistivity as a function of magnetic field at different temperatures. (c) Carrier mobility as a function of the temperature. (d) Carrier concentration as a function of the temperature. (e) Comparison of carrier concentration and mobility of typical kagome-lattice materials. Data for other typical kagome-lattice materials are from refs[20, 43].

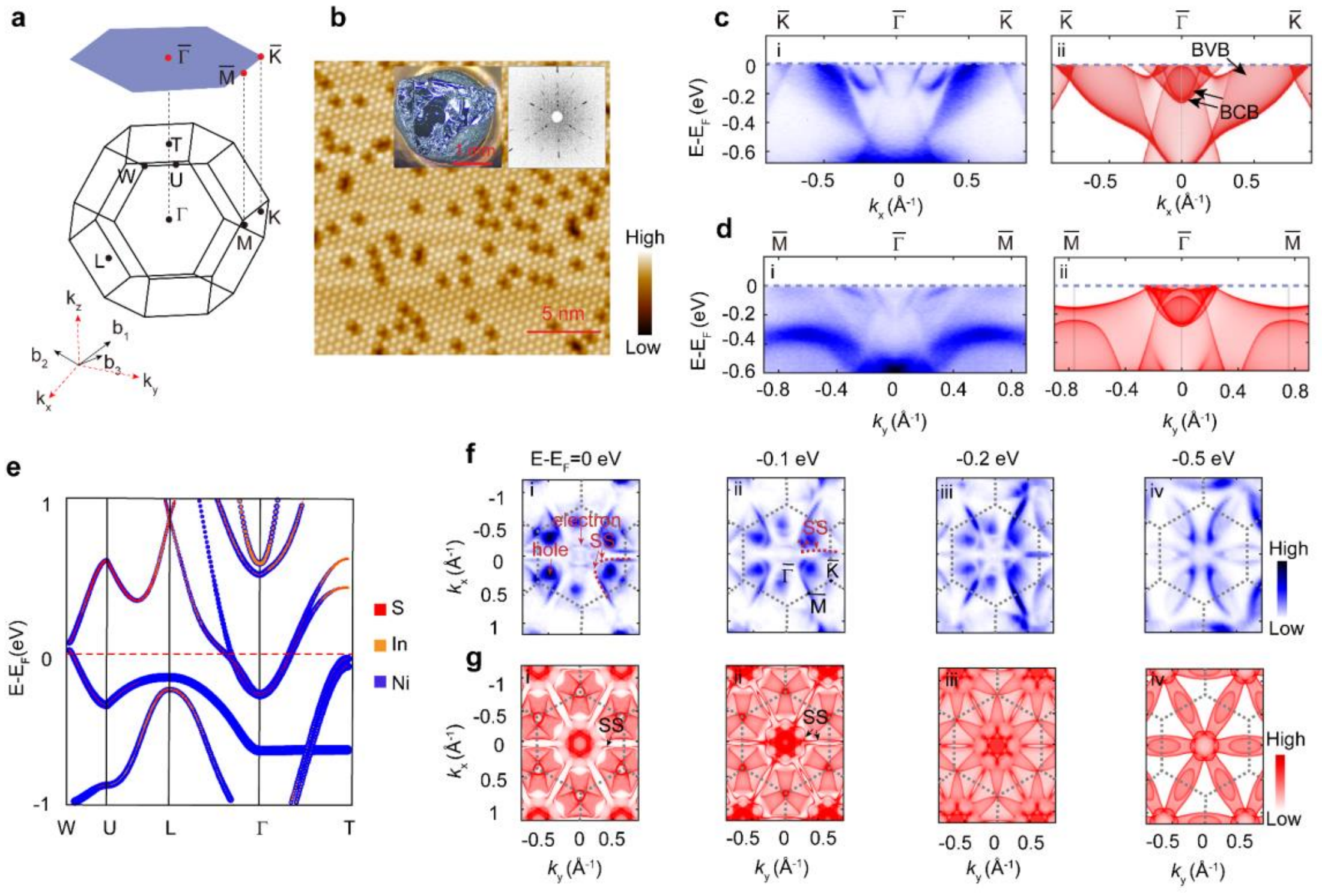

**Figure 3** The electronic structure of $Ni_3In_2S_2$ measured by ARPES. (a) The bulk BZ and its projection on the (001) surface. (b) STM topography image of the cleaved (001) surface. Left inset: cleaved surface of $Ni_3In_2S_2$ single crystal. Right inset: Laue pattern showing the high quality of the $Ni_3In_2S_2$ crystal. (c) (i-ii) High-symmetry cut along the $\bar{K}-\bar{\Gamma}-\bar{K}$ direction and corresponding calculated band dispersion. (d) (i-ii) High-symmetry cut along the $\bar{M}-\bar{\Gamma}-\bar{M}$ direction and corresponding calculated band dispersion, data are mirror symmetrized according to the crystal symmetry. (e) The calculated bulk band structure in the BZ of the primitive cell with their orbital compositions labelled in different colors. (f) (i-iv) Photoemission intensity map of constant energy contours (CECs) at 0, 0.1, 0.2 and 0.5 eV below $E_F$, respectively. (g) (i-iv) corresponding calculated CECs at the same energies. BCB: bulk conduction band; BVB: bulk valence band. SS: surface state band.

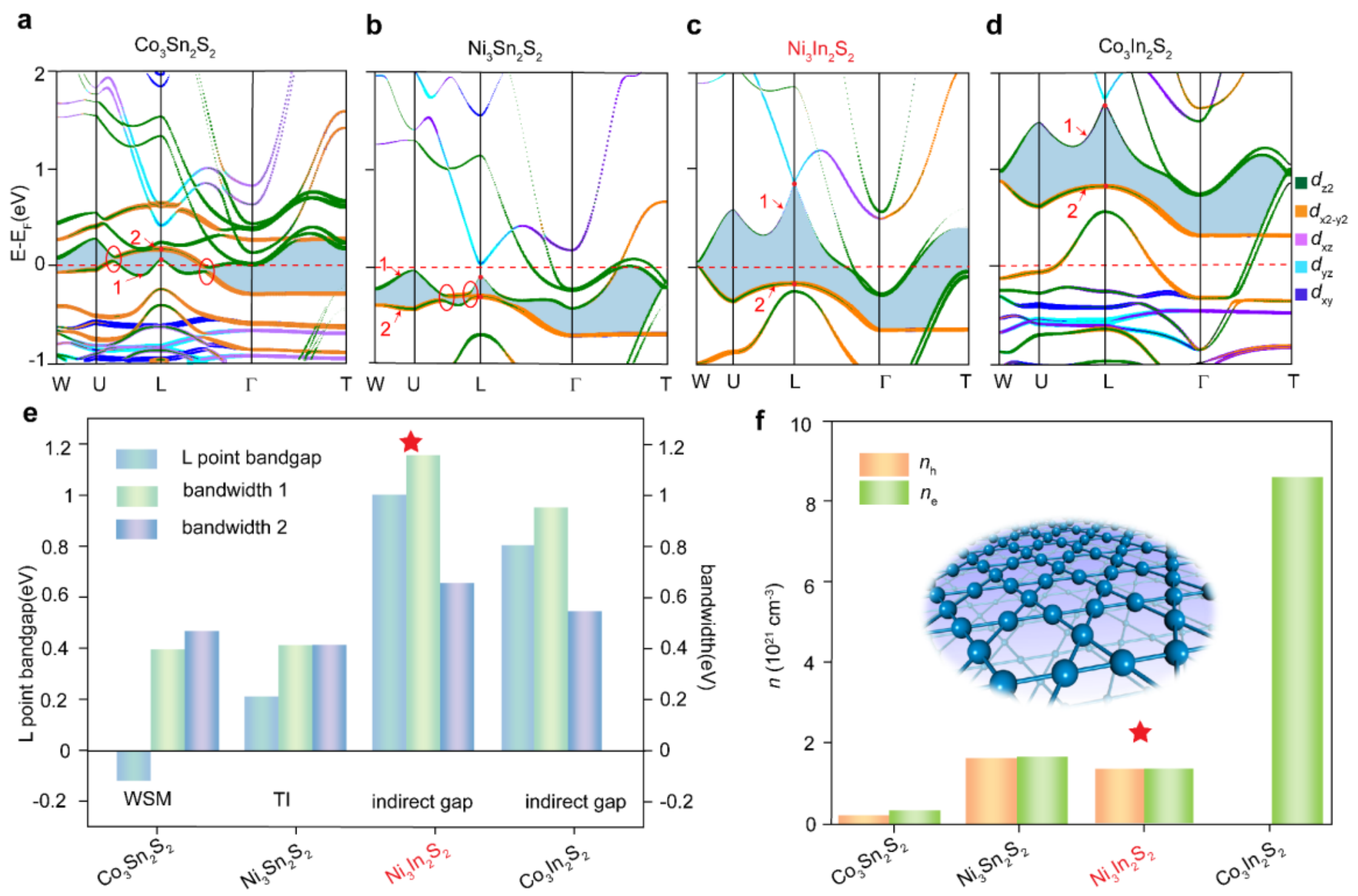

**Figure 4** Electronic origin of the transport properties in $Ni_3Sn_2S_2$. (a)-(d) Plot of the projected band structure on to Co/Ni 3*d* orbitals for $Co_3Sn_2S_2$, $Ni_3Sn_2S_2$, $Ni_3In_2S_2$ and $Co_3In_2S_2$, respectively. 1 and 2 label the bands with $dz^2$ and $dx^2$-$y^2$ orbitals, red circles indicate the band inversion points. Shaded area indicates the local band gaps between bands 1 and 2. (e) Summary of the local band gap between band 1 and 2 at L point, and bandwidth of 1 and 2 band for the four compounds. (f) Summary of the carrier concentration for the four compounds. WSM: Weyl semimetal, TI: topological insulator.

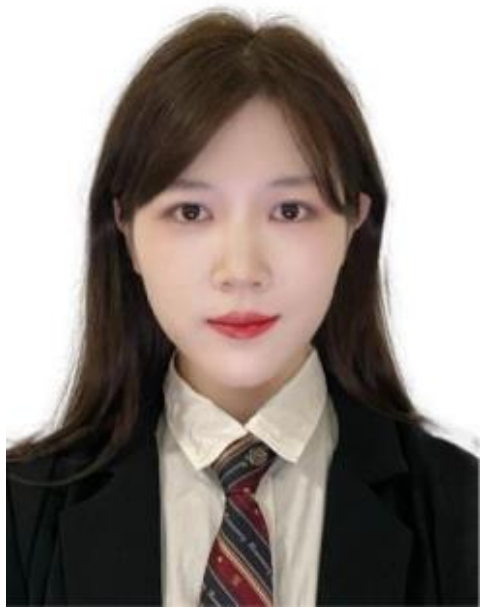

**Hongwei Fang** received her BSc degree in physics from Shandong Normal University. She is currently a PhD candidate in the ShanghaiTech University under the supervision of Prof. Zhongkai Liu. Her research interest focuses on studying the electronic properties of topological materials using angle-resolved photoemission spectroscopy (ARPES).

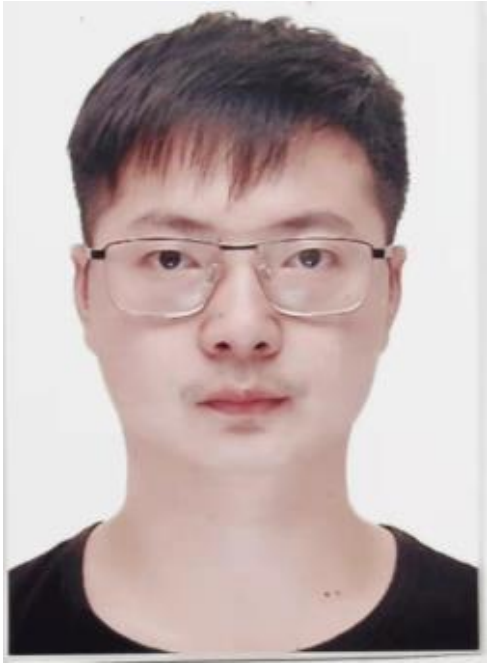

**Meng Lyu** obtained his PhD degree in the institute of physics (IOP UCAS). He is now a postdoctor in Prof. Enke Liu’s group. His research interest focuses on the exploration and study the electrical and thermal transport properties of the magnetic topological semimetals and strongly correlated electron system.

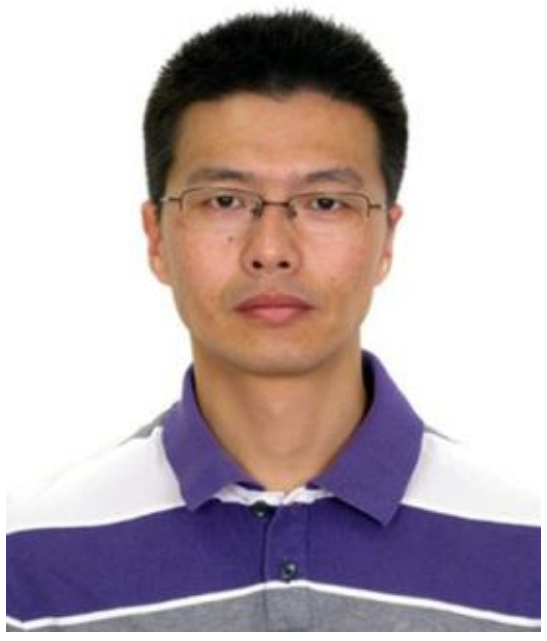

**Wujun Shi** obtained his PhD degree in Physics from Nanjing University in 2013. From 2013 to 2018, he worked as Postdoctoral Research Fellow in Tsinghua University, ShanghaiTech University and Max Planck Institute for Chemical Physics of Solids (MPI-CPfS), respectively. In 2018, he joined ShanghaiTech University, and now holds the position of Associate Researcher. His research focuses on the electronic structure of quantum materials.

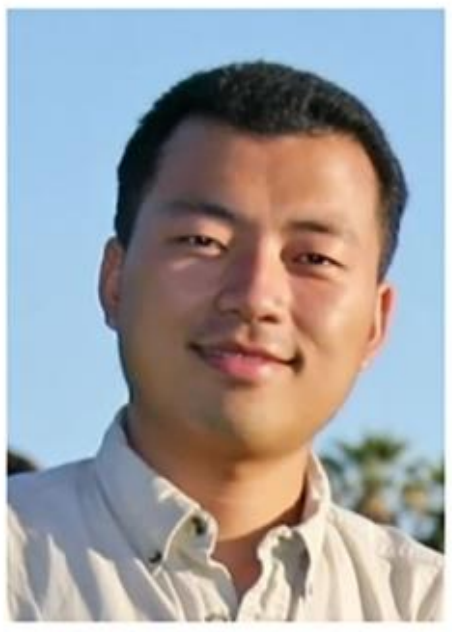

**Yulin Chen** is an associate professor at the Department of Physics, University of Oxford. His research interest focuses on the understanding and application of novel properties of quantum materials. He is also interested in developing advanced ARPES instruments with new capabilities such as spin, spatial, and time resolutions.

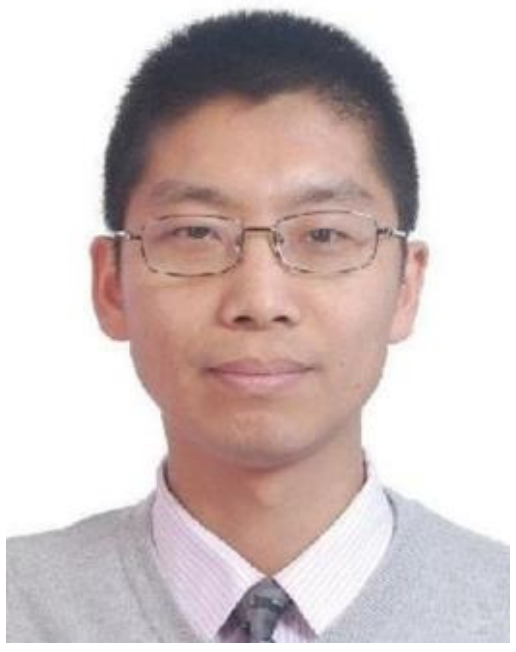

**Enke Liu** obtained his PhD degree in magnetism of condensed matter physics from Institute of Physics (IOP) Chinese Academy of Sciences in 2012. In the same year, he began his research career in IOP and now holds the position of Professor in physics. His research focuses on the new systems, novel states and electric/thermal transport of magnetic topological semimetals and magnetic phase-transformation materials.

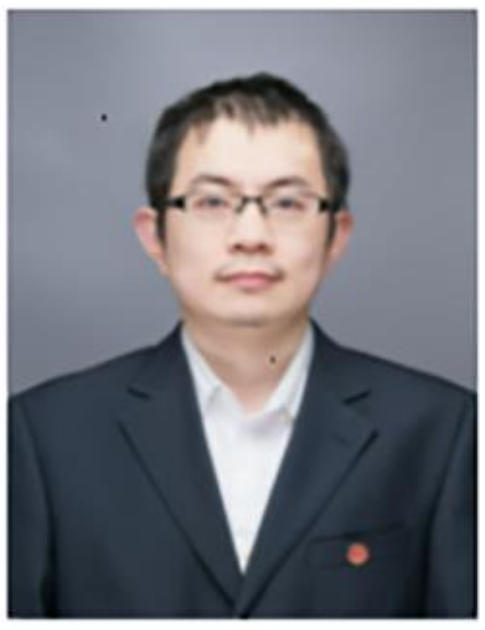

**Zhongkai Liu** received his BSc degree in Physics from Tsinghua University in 2006 and PhD degree from Stanford University in 2014. He is an associate professor at the School of Physical Science and Technology, ShanghaiTech University. His research interest focuses on characterizing the electronic structure and understanding the emergent phenomena of advanced materials with ARPES.

## kagome 晶格补偿型半金属 $Ni_3In_2S_2$ 中的创纪录的高迁移率和极大磁电阻现象

房红伟[1,2†]，吕孟[3†]，苏豪[1,2]，袁健[1,2]，李一苇[1]，徐丽璇[4]，刘帅[1,2]，魏立阳[1,2]， 刘馨琪[1,5]，杨海峰[1], 姚岐[1,5]，王美晓[1,5]，郭艳峰[1]，史武军[6,7*]，陈宇林[1,5,8*]，刘恩克[3*]，柳仲楷[1,5*]

**摘要** 具有kagome晶格的晶体具有许多有趣的性质，包括受挫磁阻、电荷有序、拓扑态、超导和关联现象。为了在电子学和自旋电子学应用中实现高性能kagome晶格化合物，需要对能带结构仔细调整。本文采用输运测量、角分辨光电子能谱和从头计算等方法研究了kagome晶格晶体$Ni_3In_2S_2$的电子结构。输运测量表明$Ni_3In_2S_2$是一种在kagome晶格材料中具有创纪录的高载流子迁移率(空穴和电子的迁移率约为8683 cm$^2$ V$^{-1}$ S$^{-1}$和7356 cm$^2$ V$^{-1}$ S$^{-1}$)和极大磁电阻(在2 K和13 T时为15518%)的补偿半金属。Ni在kagome晶格中的3d电子导致的非直接带隙、小的电子/空穴口袋和大的带宽的能带结构特征很好地解释了这些特殊的性质。这项工作表明晶体场和掺杂是优化kagome晶格晶体输运特性的关键因素。我们的工作为kagome晶格半金属在电子学和自旋电子学方面的应用提供了材料基础和优化路径。